\documentclass{article}
\usepackage[utf8]{inputenc}
\usepackage{bm}
\usepackage{amsmath}
\usepackage[numbers,square,sort&compress]{natbib} 
\usepackage{amssymb} 
\usepackage{hyperref} 
\usepackage{geometry}
\title{The Coherent State Projection Eigenvalue Theorem}
\author{Peter D. Drummond and Margaret D. Reid\\ 
Centre for Quantum Science and Technology Theory,\\
Swinburne University of Technology, Melbourne 3122, Australia\\
Correspondence: peterddrummond@protonmail.com}

\begin{document}
\maketitle

\abstract{Recent papers have developed a cyclic symmetry projected extension
for coherent state representations of bosonic Hilbert space. This
extension can greatly reduce sampling errors in phase-space simulations
of bosonic quantum systems that have symmetries. Here we prove an
eigenvalue theorem for the continuum limit of the projected coherent
states. Our intuitively appealing result is that 
in the continuum limit the projected coherent states are 
eigenstates of the symmetry generating operator. This result gives
a differential identity in phase-space that includes the conserved
eigenvalues. Detailed applications of these differential identities
to obtaining improved Fokker--Planck and stochastic equations in many-body
physics will be given elsewhere.}


\section{Introduction}

Coherent states \citep{Schrodinger_CS,Bargmann:1961,Glauber_1963_P-Rep}
are a complete basis for a bosonic Hilbert space. This allows them
to be used in many applications as a phase-space expansion \citep{drummond1980generalised,Carter:1987}.
Such methods complement traditional number-state expansions. They
can give both exact and probabilistic results for bosonic many-body
quantum systems \citep{drummond1980quantum,drummond1981nonequilibrium,Carter:1987,smith1988phase,ng2011exact,swislocki2016quantum,drummond2016quantum,stowasser2024stochastic,Najafabadi2026benchmarking}.
However, if there is any symmetry present, the coherent state basis
may be too large. For these cases it is preferable to use symmetry-projected
coherent states.

This is because when there is a dynamical conservation law, the initial
state may only support some of the conserved eigenvalues \citep{drummond2016coherent,drummond2026matrix}.
In these cases the coherent state basis includes parts of Hilbert
space that are not dynamically accessible. Under such conditions,
it is better to use projected coherent states \citep{drummond2016coherent}
to preserve the underlying symmetry. Projections reduce the Hilbert
space volume, while keeping the other properties of coherent states.
In high-dimensional, multimode systems, the effects can be extremely
dramatic. Projections can transform a simulation from one that is
impractical, owing to sampling errors, to an efficient computational
method.

This work proves an eigenvalue theorem for coherent projections on
a bosonic Hilbert space when there are continuous symmetries. Standard
coherent states and phase-space representations \citep{Glauber_1963_P-Rep}
do not include symmetries, which means that if there are conservation
laws, phase-space methods in the unprojected space can develop distribution
tails that make sampling inefficient. The example of validating a
lossless Gaussian boson sampling (GBS) quantum computer with photon-number
detection is important, since complexity theory arguments show that
quantum advantage is more likely in such lossless cases \citep{Aaronson2011,Aaronson:2014,ehrenberg2025transition,ehrenberg2025second}.
This has a conserved number parity symmetry which must be included
in phase space to allow efficient simulation \citep{dellios2026matrix,drummond2026matrix}.
Other examples of symmetry in coherent states include the use of number
conservation in simulating interacting Bose gases \citep{Carusotto:2001},
to reduce sampling errors and extend simulation times, and for generating
the Lissajous coherent states for coupled oscillators \citep{russo2026lissajous}.

In the case of GBS quantum computers, their photo-count distributions
cannot be exactly computed or sampled above about $100$ photons and
modes \citep{quesada2020exact,bulmerBoundaryQuantumAdvantage2022a}.
Experiments have now reached $8,176$ modes and $3,050$ detected
photons \citep{liu2025robust}. To validate large experiments, positive-P
simulations of grouped click distributions were demonstrated for up
to $16,384$ modes \citep{drummondSimulatingComplexNetworks2022}.
Phase-space simulations can also be used in other large Hilbert spaces
where orthogonal expansions are impractical and the states do not
factorize. Yet, when number parity is conserved in low-loss GBS experiments,
unprojected positive-P methods develop large sampling errors \citep{dellios2025validationPNR},
making it essential to use a parity projected phase-space \citep{drummond2026matrix}.
Other applications of such coherent parity projections include the
solutions of nonlinear damped systems with parametric driving \citep{Wolinsky_PRL1988,Reid_PRA1992,Munro_PRA1995}.

Our general approach uses group theory to construct the projected
coherent states from a superposition of the unprojected coherent states.
The advantage of this is that the differential identities of the original
coherent states can largely be taken over into the projected states.
This allows scalable phase-space and stochastic techniques to be used
efficiently with symmetries. Previous papers have treated either number
conservation or discrete symmetries like parity. In this work we extend
this to general continuous symmetries with commuting generators. For
such cases, an eigenvalue theorem was conjectured previously \citep{dellios2026matrix},
which we now prove. The significance of this is that it allows general
differential identities to be obtained that include the conserved
quantities associated with continuous symmetries.

\section{Cyclic projected coherent states}

We first recapitulate the notation and group setting. The starting
point of these studies is an $M$-mode bosonic Hilbert space $\mathcal{H}$,
which has a complete basis of unnormalized coherent states $\left\Vert \boldsymbol{\alpha}\right\rangle $,
where $\boldsymbol{\alpha}$ is an $M$-dimensional complex vector,
the coherent amplitude. The symmetries used are defined by a set of
$\mathcal{N}$ unitary matrix cyclic group transformations $\bm{U}_{\nu}=\exp\left(i\phi_{\nu}\bm{g}_{\nu}\right)$.
These act on the coherent amplitudes. In the following, any $\mathcal{N}$-vector
of multiple symmetries is denoted by a tilde, as in $\tilde{p}$.

The phase shift is defined as $\phi_{\nu}=2\pi/\mathcal{M}_{\nu}$
for a cyclic group of period $\mathcal{M}_{\nu}$. The group structure
is determined by the $M\times M$ Hermitian matrices $\bm{g}_{\nu}$,
which are assumed to commute with each other and to have integer eigenvalues.
As a result, they can be simultaneously diagonalized with an appropriate
unitarily transformed basis for the coherent amplitudes, giving diagonal
matrices $\bm{h}_{\nu}$, with eigenvalues $h_{k\nu}$, where $k=1,\ldots,M$
labels the modes.

The discrete symmetry matrix group elements can then be labeled 
\begin{equation}
\bm{U}^{\tilde{q}}=\exp\left(i\sum_{\nu=1}^{\mathcal{N}}q_{\nu}\phi_{\nu}\bm{g}_{\nu}\right).
\end{equation}
Given this starting point, the projected coherent states are defined
through an equivalent group of operator transformations $\hat{U}^{\tilde{q}}$
on the coherent states $\left\Vert \boldsymbol{\alpha}\right\rangle $,
such that 
\begin{equation}
\hat{U}^{\tilde{q}}=\exp\left(i\sum_{\nu=1}^{\mathcal{N}}q_{\nu}\phi_{\nu}\hat{g}_{\nu}\right),
\end{equation}
where $\hat{g}_{\nu}\equiv\hat{\bm{a}}^{\dagger}\bm{g}_{\nu}\hat{\bm{a}}$
are commuting self-adjoint operators acting on the Hilbert space.

We can now obtain the symmetry projection operator, $\hat{\Pi}_{\tilde{p}}^{\tilde{\mathcal{M}}}$,
for a vector of quantum numbers $\tilde{p}=\left[p_{1},p_{2},\ldots,p_{\mathcal{N}}\right]\in\mathbb{Z}^{\mathcal{N}}$.
This projection operator is defined as: 
\begin{equation}
\hat{\Pi}_{\tilde{p}}^{\tilde{\mathcal{M}}}\equiv\frac{1}{\mathcal{M}}\sum_{\tilde{q}}e^{i\tilde{\phi}\tilde{q}\left(\tilde{\hat{g}}-\tilde{p}\right)}.\label{eq:projector_definition}
\end{equation}
For compactness, we define $\mathcal{M}\equiv\prod_{\nu}\mathcal{M}_{\nu}$.
Any double or triple product of $\mathcal{N}$-vectors includes an
implied summation over $\mathcal{N}$, and an explicit sum over $\tilde{q}$
is used to indicate a multiple summation over all its $\mathcal{M}$
possible elements: 
\begin{equation}
\sum_{\tilde{q}}\equiv\sum_{q_{1}=0}^{\mathcal{M}_{1}-1}\sum_{q_{2}=0}^{\mathcal{M}_{2}-1}\dots\sum_{q_{\mathcal{N}}=0}^{\mathcal{M}_{\mathcal{N}}-1}.
\end{equation}

This allows one to define the projected coherent states $\left\Vert \boldsymbol{\alpha}\right\rangle _{\tilde{p}}$
as a linear superposition of the original coherent states, so that
\begin{align}
\left\Vert \boldsymbol{\alpha}\right\rangle _{\tilde{p}}^{\tilde{\mathcal{M}}} & =\hat{\Pi}_{\tilde{p}}^{\tilde{\mathcal{M}}}\left\Vert \boldsymbol{\alpha}\right\rangle \\
 & =\frac{1}{\mathcal{M}}\sum_{\tilde{q}}e^{-i\tilde{\phi}\tilde{q}\tilde{p}}\left\Vert \boldsymbol{\alpha}^{\left(\tilde{q}\right)}\right\rangle .\nonumber 
\end{align}
Here the coherent state amplitude $\boldsymbol{\alpha}^{\left(\tilde{q}\right)}\equiv\bm{U}^{\tilde{q}}\boldsymbol{\alpha}$
belongs to the orbit generated by the product of cyclic groups. Changing
$p_{\nu}$ by an integer multiple of $\mathcal{M}_{\nu}$ does not
change the resulting cyclic projector.

\section{The limiting eigenvalue theorem}

We start with two results that are already known \citep{dellios2026matrix}.
There is an eigenvalue property of the projected coherent states,
which is that every nonzero superposition state $\left\Vert \boldsymbol{\alpha}\right\rangle _{\tilde{p}}$
is an eigenstate of every $\hat{U}_{\nu}$, i.e., 
\begin{equation}
\hat{U}_{\nu}\left\Vert \boldsymbol{\alpha}\right\rangle _{\tilde{p}}^{\tilde{\mathcal{M}}}=e^{i\phi_{\nu}p_{\nu}}\left\Vert \boldsymbol{\alpha}\right\rangle _{\tilde{p}}^{\tilde{\mathcal{M}}},\qquad\nu=1,\ldots,\mathcal{N}.\label{eq:full_eigenvalue}
\end{equation}

This property can be used to generate differential identities acting
on the coherent states. However, these are generally of infinite order,
since the operators $\hat{g}_{\nu}$ which can be mapped to differential
operators acting on the coherent states, are exponentiated in the
group generator $\hat{U}$. Such identities are interesting in themselves,
but do not generate a normal, second-order Fokker--Planck equation.

For continuous Fokker--Planck equations and stochastic differential
equations, an eigenvalue of a quadratic operator expression can be
used to generate additional first- and second-order differential identities.
A useful case is when $\bm{g}_{\nu}$ is the identity matrix, so $\hat{g}$
is the number operator $\hat{n}=\sum_{k}\hat{a}_{k}^{\dagger}\hat{a}_{k}$,
and the symmetry is a phase symmetry. One can prove \citep{dellios2026matrix}
that for an index $n\ge0$, 
\begin{equation}
\lim_{\mathcal{M}\rightarrow\infty}\left[\hat{n}-n\right]\left\Vert \boldsymbol{\alpha}\right\rangle _{n}^{\mathcal{M}}=0.
\end{equation}

This can be used to generalize known Fock state stochastic equations
\citep{carusotto2001exact,carusotto2003exact,Carusotto:2001}. It
was conjectured that a similar result ought to hold for any continuous
symmetry, on the grounds that the exponential in Equation~\eqref{eq:full_eigenvalue}
can be expanded to first order in $\phi_{\nu}$ for $\phi_{\nu}\rightarrow0$.
We prove below that this is indeed the correct limiting eigenvalue
result, provided one takes account of $\mathcal{M}$-periodicity in
the cyclic projector indices. That is, the limiting eigenvalue of
$\hat{g}_{\nu}$ may be outside the usual range of $0\le p_{\nu}<\mathcal{M}_{\nu}$,
and can take negative as well as positive or zero values.

\subsection*{Limiting eigenvalue theorem}\label{thm:limiting-eigenvalue}

Let $\hat{g}_{\nu}=\hat{\boldsymbol{a}}^{\dagger}\boldsymbol{g}_{\nu}\hat{\boldsymbol{a}}$,
where $\boldsymbol{g}_{\nu}$ is a fixed finite-dimensional Hermitian
matrix with integer eigenvalues. In the single-symmetry case, write
$\hat{g}=\hat{g}_{\nu}$ and let $\hat{E}_{p}$ denote its spectral
projector for eigenvalue $p$. For fixed $p\in\mathbb{Z}$ and fixed
$\boldsymbol{\alpha}\in\mathbb{C}^{M}$, the cyclically projected
coherent states with any single projection satisfy 
\begin{equation}
\lim_{\mathcal{M}\rightarrow\infty}\left\Vert \boldsymbol{\alpha}\right\rangle _{p}^{\mathcal{M}}=\hat{E}_{p}\left\Vert \boldsymbol{\alpha}\right\rangle .\label{eq:coherent-state-limit}
\end{equation}
More generally, for fixed $\tilde{p}$ and the commuting single or
multiple projections defined above, then if $p_{\nu}$ is a fixed
eigenvalue of $\hat{g}_{\nu}$, one has that:
\begin{equation}
\lim_{\mathcal{M}_{\nu}\to\infty}\left\Vert (\hat{g}_{\nu}-p_{\nu})\left\Vert \boldsymbol{\alpha}\right\rangle _{\tilde{p}}^{\tilde{\mathcal{M}}}\right\Vert =0.
\end{equation}
The limits are in the Hilbert-space norm. If $\hat{E}_{p}\left\Vert \boldsymbol{\alpha}\right\rangle $
is nonzero, it is an eigenvector of $\hat{g}$ with eigenvalue $p$,
and the normalized single-projection states converge to its normalization. 

The cyclic order $\mathcal{M}$ and the number of retained modes $M$
are independent here. For translational symmetry, the use of Fourier
modes on a periodic domain admits continuous translations with integer
mode indices, which allows $\mathcal{M}\rightarrow\infty$, but a
finite lattice generally has only a discrete symmetry with a finite
$\mathcal{M}$. 

The theorem is proved below, initially for a single projector and
then for multiple projections.

\section{Single projector case}

We first prove this limiting result for a single projector.

\subsection{Spectral projectors}

The spectral theorem and the Hermitian generator construction are
given in general terms in Ref.~\citep{Teschl2014}, Sections 3.1
and 5.1.

Here we give a direct construction in terms of the bosonic number
state basis. If $\boldsymbol{g}=\boldsymbol{V}\boldsymbol{h}\boldsymbol{V}^{\dagger}$,
where $\boldsymbol{V}$ is the diagonalizing unitary and $\boldsymbol{h}$
is diagonal with $\hat{\boldsymbol{b}}=\boldsymbol{V}^{\dagger}\hat{\boldsymbol{a}}$,
then $\hat{g}$ can be expanded in terms of the eigenvalues $h_{k}$
of $\boldsymbol{h}$ as follows: 
\begin{equation}
\hat{g}=\sum_{k=1}^{M}h_{k}\hat{b}_{k}^{\dagger}\hat{b}_{k},\qquad h_{k}\in\mathbb{Z}.
\end{equation}
In the occupation number basis $|\boldsymbol{n}\rangle_{b}$ of the
transformed mode operators $\hat{b}_{k}$, the spectral projectors
$\hat{E}_{m}$ are defined as 
\begin{equation}
\hat{E}_{m}=\sum_{\substack{\boldsymbol{n}\in\mathbb{N}_{0}^{M}\\
\sum_{k}h_{k}n_{k}=m
}
}|\boldsymbol{n}\rangle_{b}\,{}_{b}\langle\boldsymbol{n}|.\label{eq:spectral_projector_definition}
\end{equation}

All number states correspond to a spectral projector with some index
$m$, and a projector $\hat{E}_{m}$ with no corresponding number
state is zero. These projectors obey the usual properties that: 
\begin{align}
\hat{E}_{m}^{\dagger} & =\hat{E}_{m},\nonumber \\
\hat{E}_{m}\hat{E}_{n} & =\delta_{mn}\hat{E}_{m},\nonumber \\
\sum_{m\in\mathbb{Z}}\hat{E}_{m} & =\hat{I}.\label{eq:Spectral-properties}
\end{align}

The identity sum means that, after application to any fixed $\left|\psi\right\rangle \in\mathcal{H}$,
its partial sums converge in Hilbert-space norm. In particular, one
obtains that: 
\begin{align}
\left|\psi\right\rangle  & =\sum_{m\in\mathbb{Z}}\hat{E}_{m}\left|\psi\right\rangle ,\nonumber \\
\sum_{m\in\mathbb{Z}}\|\hat{E}_{m}\left|\psi\right\rangle \|^{2} & =\|\left|\psi\right\rangle \|^{2}.\label{eq:projection_sum}
\end{align}
For the unbounded generator $\hat{g}$, its domain is 
\begin{equation}
D(\hat{g})=\left\{ \left|\psi\right\rangle \in\mathcal{H}:\sum_{m\in\mathbb{Z}}m^{2}\|\hat{E}_{m}\left|\psi\right\rangle \|^{2}<\infty\right\} .\label{eq:generator_domain}
\end{equation}
Substituting the allowed values of $n_{k}$ given in the definition,
Equation~\eqref{eq:spectral_projector_definition}, its action on
any state $|\psi\rangle\in D(\hat{g})$ can be written as: 
\begin{align}
\hat{g}\left|\psi\right\rangle  & =\sum_{m\in\mathbb{Z}}m\hat{E}_{m}\left|\psi\right\rangle ,\label{eq:g-psi-spectral-result}
\end{align}
and the corresponding norm is: 
\begin{equation}
\|\hat{g}\left|\psi\right\rangle \|^{2}=\sum_{m\in\mathbb{Z}}m^{2}\|\hat{E}_{m}\left|\psi\right\rangle \|^{2}.
\end{equation}
Theorem 3.6 of \citep{Teschl2014} has a general spectral theorem
while Eq. (3.26) gives the domain condition. In particular, coherent
states lie in the domain. Using the spectral projector product rule
in Equation~\eqref{eq:Spectral-properties}, one immediately obtains
an eigenvalue result for the spectral projector, namely that if $p$
is an eigenvalue
\begin{equation}
\hat{g}\hat{E}_{p}\left\Vert \boldsymbol{\alpha}\right\rangle =p\hat{E}_{p}\left\Vert \boldsymbol{\alpha}\right\rangle .\label{eq:spectral_eigenvalue}
\end{equation}

\subsection{Cyclic and spectral projectors}

For a positive integer $\mathcal{M}$ and an integer $p$, define
the cyclic projector as above: 
\begin{equation}
\hat{\Pi}_{p}^{(\mathcal{M})}=\frac{1}{\mathcal{M}}\sum_{q=0}^{\mathcal{M}-1}\exp\left[\frac{2\pi iq}{\mathcal{M}}(\hat{g}-p\hat{I})\right].
\end{equation}
We keep an integer representative $p$ fixed when taking the limit
of large $\mathcal{M}$. Next, one can use the expansion of the identity
to give that: 
\begin{equation}
\hat{\Pi}_{p}^{(\mathcal{M})}=\frac{1}{\mathcal{M}}\sum_{q=0}^{\mathcal{M}-1}\sum_{m\in\mathbb{Z}}\exp\left[\frac{2\pi iq}{\mathcal{M}}(\hat{g}-p\hat{I})\right]\hat{E}_{m}.\label{eq:Expansion_of_cyclic_projector}
\end{equation}

Using Equation~\eqref{eq:g-psi-spectral-result} together with mutual
orthogonality of the projectors gives, for every $m$, 
\begin{equation}
\exp\left[\frac{2\pi iq}{\mathcal{M}}(\hat{g}-p\hat{I})\right]\hat{E}_{m}=\exp\left[\frac{2\pi iq}{\mathcal{M}}(m-p)\right]\hat{E}_{m}.
\end{equation}
Next, we introduce the periodic Kronecker delta function, 
\begin{equation}
\delta_{r,0}^{(\mathcal{M})}=\frac{1}{\mathcal{M}}\sum_{q=0}^{\mathcal{M}-1}\exp\left(\frac{2\pi iqr}{\mathcal{M}}\right),\qquad r\in\mathbb{Z}.
\end{equation}

If $r$ is a multiple of $\mathcal{M}$, every term is unity and $\delta_{r,0}^{(\mathcal{M})}=1$.
Otherwise the finite geometric-series identity gives $\delta_{r,0}^{(\mathcal{M})}=0$.
Hence, one sees immediately from Equation~\eqref{eq:Expansion_of_cyclic_projector}
that 
\begin{equation}
\hat{\Pi}_{p}^{(\mathcal{M})}=\sum_{m\in\mathbb{Z}}\delta_{m-p,0}^{(\mathcal{M})}\hat{E}_{m}.
\end{equation}
This gives an expansion of the cyclic projector as a series of spectral
projectors: 
\begin{equation}
\hat{\Pi}_{p}^{(\mathcal{M})}=\sum_{\ell\in\mathbb{Z}}\hat{E}_{p+\ell\mathcal{M}}.\label{eq:spectral_projector_identity}
\end{equation}

\subsection{The large-order limit}

\label{sec:large-order-limit}

The large-order limit is defined by $\mathcal{M}\longrightarrow\infty$
through positive integers, with $\hat{g},\ p,\ \left|\psi\right\rangle $
all fixed. For the operators and each fixed vector $\left|\psi\right\rangle $,
it gives convergence in Hilbert-space norm, which is not generally
an operator-norm convergence. Subtracting the term with $\ell=0$
from the spectral projector identity, Equation~\eqref{eq:spectral_projector_identity}:
\begin{equation}
\left[\hat{\Pi}_{p}^{(\mathcal{M})}-\hat{E}_{p}\right]\left|\psi\right\rangle =\sum_{\ell\ne0}\hat{E}_{p+\ell\mathcal{M}}\left|\psi\right\rangle .
\end{equation}
Different terms lie in orthogonal eigenspaces. Therefore, 
\begin{equation}
\Delta\equiv\left\Vert \left[\hat{\Pi}_{p}^{(\mathcal{M})}-\hat{E}_{p}\right]\left|\psi\right\rangle \right\Vert ^{2}=\sum_{\ell\ne0}\left\Vert \hat{E}_{p+\ell\mathcal{M}}\left|\psi\right\rangle \right\Vert ^{2}.
\end{equation}

For every $\ell\ne0$, the following inequalities hold: 
\begin{equation}
|p+\ell\mathcal{M}|\ge|\ell|\mathcal{M}-|p|\ge\mathcal{M}-|p|.
\end{equation}
For large enough $\mathcal{M}$, since $p$is fixed, one can take
$\mathcal{M}>|p|$, so
\begin{equation}
0\le\Delta\le\sum_{|m|\ge\mathcal{M}-|p|}\|\hat{E}_{m}\left|\psi\right\rangle \|^{2}.
\end{equation}

The upper bound is the tail of the convergent nonnegative projection
series given in Equation~\eqref{eq:projection_sum}, which vanishes
for large $\mathcal{M}$. This proves that there is convergence in
the sense that: 
\begin{equation}
\lim_{\mathcal{M}\rightarrow\infty}\hat{\Pi}_{p}^{(\mathcal{M})}\left|\psi\right\rangle =\hat{E}_{p}\left|\psi\right\rangle .\label{eq:Large-M-limit}
\end{equation}
While operator-norm convergence need not hold, this stronger convergence
is not needed. Note that $p$ is not restricted to positive integer
values, and can be any integer eigenvalue of the generator $\hat{g}$.

\subsection{The cyclic eigenvalue}

We define $\left|\psi\right\rangle $ to be 
\begin{equation}
\left|\psi\right\rangle =\left[\hat{g}-p\right]\left\Vert \boldsymbol{\alpha}\right\rangle .
\end{equation}
We now show that $(\hat{g}-p)\|\boldsymbol{\alpha}\rangle$ has finite
norm. The strong convergence of the cyclic projectors can then be
applied to this fixed vector. Let $\boldsymbol{\beta}=\boldsymbol{V}^{\dagger}\boldsymbol{\alpha},$
where $\boldsymbol{V}$ is the unitary that diagonalizes $\bm{g},$
and define 
\begin{align}
s & =\sum_{k}|\beta_{k}|^{2},\nonumber \\
\mu & =\sum_{k}h_{k}|\beta_{k}|^{2},\nonumber \\
v & =\sum_{k}h_{k}^{2}|\beta_{k}|^{2}.
\end{align}
The unnormalized coherent state has the occupation-number expansion
\begin{equation}
\left\Vert \boldsymbol{\alpha}\right\rangle =\sum_{\boldsymbol{n}\in\mathbb{N}_{0}^{M}}\frac{\prod_{k}\beta_{k}^{n_{k}}}{\sqrt{\prod_{k}n_{k}!}}|\boldsymbol{n}\rangle_{b}.
\end{equation}
Consequently, $\left\Vert (\hat{g}-p)\left\Vert \boldsymbol{\alpha}\right\rangle \right\Vert ^{2}=e^{s}\bigl[(\mu-p)^{2}+v\bigr]<\infty.$
This follows by summing the first and second moments of the independent
Poisson weights, and hence the expression is finite for every coherent
state. 

The spectral expansion of the cyclic projector shows that it preserves
the generator domain, $D(\hat{g})$, defined in Eq. (\ref{eq:generator_domain}).
For any $\left|\chi\right\rangle \in D(\hat{g})$, 
\[
\sum_{m\in\mathbb{Z}}m^{2}\left\Vert \hat{E}_{m}\hat{\Pi}_{p}^{(\mathcal{M})}\left|\chi\right\rangle \right\Vert ^{2}\leq\sum_{m\in\mathbb{Z}}m^{2}\|\hat{E}_{m}\left|\chi\right\rangle \|^{2}<\infty.
\]
Hence, since these projection operators have a common set of number-state
eigenvectors, $\hat{g}\hat{\Pi}_{p}^{(\mathcal{M})}\left|\chi\right\rangle =\hat{\Pi}_{p}^{(\mathcal{M})}\hat{g}\left|\chi\right\rangle $
on this domain. In summary, $\|\boldsymbol{\alpha}\rangle\in D(\hat{g})$,
and $|\psi\rangle=(\hat{g}-p)\|\boldsymbol{\alpha}\rangle$ is a vector
in $\mathcal{H}$. The cyclic projectors preserve $D(\hat{g})$ and
commute with $\hat{g}$ on this domain. 

It therefore follows that 
\begin{equation}
\hat{\Pi}_{p}^{(\mathcal{M})}\left|\psi\right\rangle =\left[\hat{g}-p\right]\left\Vert \boldsymbol{\alpha}\right\rangle _{p}^{\mathcal{M}}.
\end{equation}
From the eigenvalue property of Equation~\eqref{eq:spectral_eigenvalue},
\begin{align}
\hat{E}_{p}\left|\psi\right\rangle  & =\left[\hat{g}-p\right]\hat{E}_{p}\left\Vert \boldsymbol{\alpha}\right\rangle =0.
\end{align}
The moment calculation shows that $|\psi\rangle\in\mathcal{H}$, so
the strong convergence limit holds. Using the limiting property in
Equation~\eqref{eq:Large-M-limit}, one obtains the required limiting
eigenvalue result: 
\begin{equation}
\lim_{\mathcal{M}\rightarrow\infty}\left[\hat{g}-p\right]\left\Vert \boldsymbol{\alpha}\right\rangle _{p}^{\mathcal{M}}=0.
\label{eq:cyclic-eigenvalue-limit}
\end{equation}

\section{Multiple projectors and differential identities}

\label{several-commuting-symmetries-and-the-scope-of-the-limit}

\subsection{Multiple projection case}

The general case of multiple generators requires that their spectral
projectors commute. In the finite-mode construction given here, the
common diagonalization of the one-particle matrices provides a common
number basis. Hence the commutation requirement is satisfied.

For this more general case, define a multiple symmetry spectral projector
as 
\begin{equation}
\hat{E}_{\tilde{p}}=\prod_{\nu=1}^{\mathcal{N}}\hat{E}_{\nu,p_{\nu}}.
\end{equation}
Recursively multiplying the one-symmetry result given above then gives:
\begin{equation}
\hat{\Pi}_{\tilde{p}}^{\tilde{\mathcal{M}}}=\sum_{\ell_{1},\ldots,\ell_{\mathcal{N}}\in\mathbb{Z}}\hat{E}_{p_{1}+\ell_{1}\mathcal{M}_{1},\ldots,p_{\mathcal{N}}+\ell_{\mathcal{N}}\mathcal{M}_{\mathcal{N}}}.
\end{equation}

With $\tilde{p}$, the generators, and $\psi$ fixed, and the $\mathcal{N}$
symmetries each taking the limit of $\mathcal{M}_{\nu}\to\infty$,
the same orthogonal-tail argument gives 
\begin{equation}
\hat{\Pi}_{\tilde{p}}^{\tilde{\mathcal{M}}}\left|\psi\right\rangle \longrightarrow\hat{E}_{\tilde{p}}\left|\psi\right\rangle .
\end{equation}
For $\left|\psi\right\rangle \in\bigcap_{\nu=1}^{\mathcal{N}}D(\hat{g}_{\nu})$
it also gives, for each $\nu$, that: 
\begin{equation}
\hat{g}_{\nu}\hat{\Pi}_{\tilde{p}}^{\tilde{\mathcal{M}}}\left|\psi\right\rangle \longrightarrow p_{\nu}\hat{E}_{\tilde{p}}\left|\psi\right\rangle .
\end{equation}
Specializing to the case of coherent states, one obtains the multiple
symmetry result: 
\begin{equation}
\hat{g}_{\nu}\left\Vert \bm{\alpha}\right\rangle _{\tilde{p}}^{\tilde{\mathcal{M}}}\longrightarrow p_{\nu}\hat{E}_{\tilde{p}}\left\Vert \bm{\alpha}\right\rangle .
\end{equation}

Here we repeat the result for a single projector to obtain the limiting
eigenvalue result for each continuous symmetry. If only some of the
symmetries need to be taken in the continuum limit, then only those
that are continuous satisfy the theorem, that is, if there is a subset
$J$ of the cyclic orders tending to infinity, then the limiting projector
is: 
\begin{equation}
\hat{\Pi}_{\tilde{p}}^{\left(\lim\right)}=\left(\prod_{\nu\in J}\hat{E}_{\nu,p_{\nu}}\right)\left(\prod_{\mu\notin J}\hat{\Pi}_{\mu,p_{\mu}}^{(\mathcal{M}_{\mu})}\right).
\end{equation}

\subsection{Eigenvalue differential identities}

We now outline how these results can be used to obtain differential
identities. It is possible to map master equation or Schrödinger equation
operators into differential identities without using the eigenvalues.
While this is useful, if there are no other mappings, it can generate
an inefficient stochastic mapping with a rapidly growing sampling
error. To overcome this, it is better to include the conserved quantities
in the dynamics.

A typical application is to use a matrix P-representation \citep{dellios2026matrix,drummond2026matrix}.
This expands the quantum density matrix $\hat{\rho}$ as: 
\begin{equation}
\hat{\rho}=\int P\left(\vec{\lambda},t\right)\hat{\Lambda}_{\mathcal{M}}\left(\vec{\lambda}\right)d\vec{\lambda}.
\end{equation}
In the case of a doubled phase-space, like the positive-P representation,
$\vec{\lambda}=\left(\bm{\alpha},\bm{\beta}^{*},\underline{\Omega}\right)$
denotes the phase-space coordinates, comprising coherent amplitudes
$\bm{\alpha}$, $\bm{\beta}$ and the $\mathcal{M}\times\mathcal{M}$
symmetry matrix $\underline{\Omega}$. More generally, one can define
$\vec{\lambda}=\left(\vec{\alpha},\underline{\Omega}\right)$, with
$C$ coherent components, $\vec{\alpha}=\left[\bm{\alpha}^{(1)},\bm{\alpha}^{(2)},\ldots\bm{\alpha}^{(C)}\right]$.

Different normalizations of $\hat{\Lambda}_{\mathcal{M}}$ are useful
in different physical applications. For the simplest case of an unnormalized
kernel $\hat{\Lambda}_{\mathcal{M}}$, one defines: 
\begin{equation}
\hat{\Lambda}_{\mathcal{M}}\left(\vec{\lambda}\right)=\operatorname{Tr}_{\mathcal{M}}\left[\underline{\Omega}\underline{\hat{K}}\left(\vec{\alpha}\right)\right],
\end{equation}
where $\operatorname{Tr}_{\mathcal{M}}$ is a trace over $\mathcal{M}\times\mathcal{M}$
matrices, and the kernel elements are defined by: 
\begin{equation}
\hat{K}_{\tilde{p}\tilde{q}}\left(\boldsymbol{\alpha},\boldsymbol{\beta}\right)=\sum_{c\neq d}\left\Vert \boldsymbol{\alpha}^{(c)}\right\rangle _{\tilde{p}}^{(\tilde{\mathcal{M}})}\left\langle \boldsymbol{\alpha}^{(d)}\right\Vert _{\tilde{q}}^{(\tilde{\mathcal{M}})}.
\end{equation}
In all such phase-space representations, to obtain equations of motion
in the phase-space, one must map quantum operators into differential
operators. The eigenvalue theorem allows one to derive additional
differential identities. These are then used to optimize the equations
of motion so that conserved quantities in the Hilbert space lead to
corresponding (nearly) conserved stochastic equations on manifolds
\citep{joseph2023midpoint}. This is a generalization of the stochastic
Fock equation approach \citep{Carusotto:2001,carusotto2001exact,Carusotto2003,carusotto2003exact},
which conserves the total boson number.

To demonstrate how to obtain the additional identities when all the
symmetries are continuous, we summarize results that were presented
elsewhere, using the eigenvalue conjecture \citep{dellios2026matrix}
that is now proved in this paper. First, define 
\begin{equation}
\left\Vert \boldsymbol{\alpha}\right\rangle _{\tilde{p}}^{(\infty)}=\hat{E}_{\tilde{p}}\left\Vert \boldsymbol{\alpha}\right\rangle .
\end{equation}
A standard result for an unnormalized coherent state is that: 
\begin{equation}
\hat{g}_{\nu}\left\Vert \boldsymbol{\alpha}\right\rangle =\sum_{j,k}g_{\nu,jk}\alpha_{k}\frac{\partial}{\partial\alpha_{j}}\left\Vert \boldsymbol{\alpha}\right\rangle .
\end{equation}

Since $\hat{E}_{\tilde{p}}$ is independent of $\boldsymbol{\alpha}$
and commutes with each $\hat{g}_{\nu}$, it follows that: 
\begin{equation}
\left[\sum_{j,k}g_{\nu,jk}\alpha_{k}\frac{\partial}{\partial\alpha_{j}}-p_{\nu}\right]\left\Vert \boldsymbol{\alpha}\right\rangle _{\tilde{p}}^{(\infty)}=0.
\end{equation}
This derives the differential identity from the limiting projected
state. Similarly, one can obtain a second-order term on differentiating
the identities with a gauge structure constant or function, $u_{\nu,i}$,
so that
\begin{equation}
u_{\nu,i}\left[\sum_{j,k}g_{\nu,jk}\alpha_{k}\frac{\partial^{2}}{\partial\alpha_{i}\partial\alpha_{j}}+\sum_{j}g_{\nu,ji}\frac{\partial}{\partial\alpha_{j}}-p_{\nu}\frac{\partial}{\partial\alpha_{i}}\right]\left\Vert \boldsymbol{\alpha}\right\rangle _{\tilde{p}}^{(\infty)}=0.\label{eq:second-order-identity}
\end{equation}

After partial integration, the above result leads to both additional
drift terms and diffusion terms in the corresponding Fokker-Planck
equations. There is a complementary identity for the $\bm{\beta}$
variable, or more generally, for each of the other phase-space components.
These can be used to modify the resulting phase-space distributions
to reduce distribution tails. Details and examples of how these identities
can be used to stabilize the stochastic simulations for minimizing
sampling errors will be treated elsewhere.

\section{Conclusions}

We have presented a proof of a conjecture about the limiting behavior
of projected coherent states constructed using Fourier transforms
of superpositions. These allow a large class of discrete symmetry
projected coherent states to be obtained, together with relevant differential
identities used in their applications. The identities can be used
to derive Fokker-Planck and stochastic differential equations for
dynamical or canonical results and numerical simulations. The result
obtained here is that in the continuum limit the projected coherent
states become eigenstates of the relevant generator. Thus, in the
case of phase symmetry one obtains an eigenstate of total particle
number, and with spatial translational symmetry one obtains an eigenstate
of total momentum. These are not single-particle eigenstates, but
include all possible multiparticle excitations and superpositions
that satisfy the eigenvalue condition.

The benefit of this approach is that the starting point is a quantum
coherent state. This means that one can readily obtain phase-space
differential identities. The coherent identities allow one to apply
scalable methods of stochastic differential equations, but for unitary
evolution without damping, sampling errors often grow in time. The
projected eigenvalue identities extends the number of differential
identities available. These can then be used to optimize phase-space
dynamics and reduce sampling errors. Further applications will be
treated elsewhere.



\subsection*{Acknowledgements}
{This publication was made possible through support from Grant 62843
of the John Templeton Foundation. The opinions expressed in this publication
are those of the authors and do not necessarily reflect the views
of the John Templeton Foundation.}








\end{document}